\documentstyle[12pt,boxedminipage,epsfig,wrapfig,floatflt,shadow]{article}

\begin{document}

\title{Improving Students' Understanding of Quantum Mechanics via the Stern-Gerlach Experiment}

\author{Guangtian Zhu and Chandralekha Singh \\ Department of Physics and Astronomy\\ University of Pittsburgh, Pittsburgh, PA, 15260}
\date{ }

\maketitle

\begin{abstract}
The Stern-Gerlach experiment has played a central role in the discovery of spin angular momentum. It 
can also play a pivotal role in teaching the formalism of quantum mechanics using a concrete example involving a finite-dimensional
Hilbert space. Using this context, students can learn about how to prepare a specific quantum state starting from an arbitrary state, issues related to
the time evolution of the wave function, 
and quantum measurement. It can also be exploited to teach students about the distinction between the physical
space where one performs the experiment and the Hilbert space where the state of the system lies and how the information about 
the state of the system in the Hilbert space can be exploited to interpret the possible outcomes of the experiment in the physical space. 
Students can learn the advantages of choosing an appropriate basis to make suitable predictions 
about the outcomes of experiments with different arrangements of Stern-Gerlach devices. 
This experiment can also help students understand that an ensemble of identically prepared systems, e.g., one in a linear superposition of 
two stationary states, is not the same as a mixture, e.g., in which half of the systems are in one stationary state and the other
half are in the other stationary state. 
Here, we discuss investigation of students' difficulties about the Stern-Gerlach experiment
by giving written tests and interviewing advanced undergraduate and graduate students in quantum mechanics courses.
We also discuss preliminary data from two quantum mechanics courses that suggest that a Quantum Interactive Learning Tutorial (QuILT) 
related to the Stern-Gerlach experiment is helpful in improving students' understanding of these concepts.
\end{abstract}

\section{Introduction}
\vspace*{-.08in}

Learning quantum mechanics is challenging~\cite{styer,zollman2,johnston,ireson,zollman,wittman,redish,my2,lincoln,categorization,mason}. 
Investigation of students' difficulties in learning quantum mechanics is a stepping stone to developing
strategies to improve student understanding~\cite{zollman2,johnston,zollman,wittman,my2,narst,hiller,chandra1,chandra2}.
In this paper, our goal is to discuss two related issues: (1) investigation of students' difficulties related to the Stern-Gerlach experiment (SGE) and (2) the development and 
evaluation of a research-based Quantum Interactive Learning Tutorial (QuILT)~\cite{singh1,quilt,grad,my3,guangtian} that strives to help students learn about 
foundational issues in quantum mechanics via the Stern-Gerlach experiment.~\cite{schroeder,styer2,stern,feynman,stern3,stern2}

In the SGE, a particle with mass, spin and/or orbital angular momentum (a particle with a magnetic dipole moment) is sent through a 
Stern-Gerlach Apparatus (SGA) with a non-uniform magnetic field. With an appropriate gradient of the external magnetic field, different components of 
the angular momentum in the
wave function can be spatially separated by coupling them with different linear momenta. By using suitable measurement devices (e.g., detectors
at appropriate locations in the path of the beam), we can use the SGE to prepare a quantum state which is different from the initial state
before the particle entered the SGA. 
The knowledge deficiencies related to the SGE discussed in the next section can be broadly divided into three levels with increased difficulty in overcoming it:
(I) lack of knowledge of relevant concepts,
(II) knowledge that cannot be interpreted correctly,
(III) knowledge that is interpreted at the basic level but cannot be used to draw inferences in specific situations.~\cite{singh1}

The SGE QuILT is based upon research on students' difficulties in learning quantum mechanics.
It strives to build on students' prior knowledge, actively engages them in the learning process and helps them build links between 
the abstract formalism and conceptual aspects of quantum physics without compromising the technical content.
The QuILT uses a guided inquiry method of learning and the various sections build on what the students did in the previous sections to
help them develop a robust knowledge structure.
As students progress through the QuILT, they first make predictions about what should happen in various situations and then they are given guidance and support
to reason through the situations appropriately and assimmilate and accommodate productive ideas into their knowledge structure.~\cite{tutor}
The SGE QuILT creates an active learning environment in which students will directly confront their misconceptions. At various stages of concept development,
the SGE QuILT often exploits computer-based visualization tools. Often these tools cause a cognitive conflict if students' initial prediction and their observations do not match.
In that case, students themselves realize that there is some inconsistency in their reasoning. Then, providing students appropriate guidance and support via the guided inquiry
approach used in the QuILT is an effective strategy to help them build 
a robust knowledge structure.

\section{Investigation of Students' difficulties}
\vspace*{-.08in}

The investigation of difficulties was carried out by administering written surveys to more than two hundred physics graduate 
students and advanced undergraduate students enrolled in quantum mechanics courses and
by conducting individual interviews with a subset of students.
The individual interviews were carried out using a think-aloud protocol~\cite{chi}
to better understand the rationale for their responses before, during and after the development of different versions of the SGE QuILT and the
corresponding pre-test and post-test.
During the semi-structured interviews, students were asked to verbalize their thought processes while they answered
questions either as separate questions before the preliminary version of the QuILT was developed or as a part of the QuILT.
Students were not interrupted unless they remained quiet for a while. 
In the end, we asked them for clarification of the issues they had not made clear earlier. 
Some of these interviews involved asking students to predict what should happen in a particular situation, having them
observe what happens in a simulation, and asking them to reconcile the differences between their prediction and observation.
After each individual interview with a particular version of the QuILT (along with the pre-test and post-test administered), 
modifications were made based upon the feedback obtained from students' performance on the QuILT (if students got stuck at a particular
point and could not make progress from one question to the next with the hints already provided, suitable modifications were made), the pre-test and the post-test. 

\vspace*{-.15in}
\subsection{Difficulty in Distinguishing between the Physical Space and Hilbert Space}
\vspace*{-.05in}

Using quantum theory, one can interpret the outcome of experiments performed, e.g., in three dimensional (3D) laboratory or physical space by
making connection with an abstract Hilbert space (state space) in which the state of the quantum system or wavefunction lies. The
physical observables that one measures in the laboratory correspond to Hermitian operators in the Hilbert space whose
eigenstates span the Hilbert space. Knowing the initial wavefunction and the Hamiltonian of the system allows
one to determine the time-evolution of the wavefunction unambiguously and the measurement postulate can be used to
determine the possible outcomes of individual measurements of an observable and their ensemble averages (expectation values).

It is difficult for many students to distinguish between vectors in the 3D laboratory space and states in Hilbert space. For example,
$S_x$, $S_y$ and $S_z$ denote the orthogonal components of the spin angular momentum vector of an electron in the 3D space, each
of which is a physical observable that can be measured in the laboratory. However, the Hilbert space corresponding to the
spin degree of freedom for a spin-1/2 particle is two dimensional (2D). In this Hilbert space, $\hat S_x$, $\hat S_y$ and $\hat S_z$
are operators whose eigenstates span the 2D space.
Thus, the eigenstates of $\hat S_x$ are vectors which span the 2D space and are orthogonal to each other
(but not orthogonal to the eigenstates of $\hat S_y$ or $\hat S_z$).
If the electron is in a magnetic field with the field gradient
in the $z$ direction in the laboratory (3D space) as in the Stern-Gerlach experiment, the magnetic field is a vector field
in the 3D space but not in the 2D Hilbert space. It does not make sense to compare vectors in the 3D space with
the vectors in the 2D space as in statements such as ``the magnetic field gradient is perpendicular to the eigenstates of $\hat S_x$".
In fact, even $L=1$ orbital angular momentum states, which are vectors in a 3D Hilbert space, are different from the 3D laboratory space. 
Unfortunately, these distinctions are difficult for students to make and such difficulties were frequently observed in response
to the survey questions and during the individual interviews. These difficulties are discussed below in the context of
the Stern-Gerlach experiment.

\vspace*{-.25in}
\subsection{Difficulty in Determining the Pattern on the Screen with Particles in Different Spin States Passing through an SGA}
\vspace*{-.09in}

Two questions we have asked the first year physics graduate students and advanced undergraduate students for several years related to the SGE 
in written tests and interviews are questions (1) and (2) in the Appendix. In one version of these questions, neutral silver atoms were replaced with
electrons and students were told to ignore the Lorentz force on the electron. 

In question (1) in the Appendix, students have to realize that the magnetic field gradient in the $-z$ direction would impart 
a spin-dependent momentum to the particle and one should observe two spots on the phosphor screen owing to the splitting
of the beam along the $z$ direction due to the particle's spin components corresponding to the
$\vert \uparrow \rangle_z$ and $\vert \downarrow \rangle_z$ states.
All responses in which students noted that there will be a splitting along the $z$ direction
were considered correct even if they did not explain their reasoning.
Only $41\%$ of the more than 200 graduate students from different universities
enrolled in a quantum mechanics course provided the correct response. These students were
given this question as a part of a survey at the beginning of graduate level quantum mechanics instruction.
Many students thought that there will only be a single spot on the phosphor screen.
During the interviews conducted with a subset of students, they were often confused about the origin of the spin-dependent momentum imparted to the particle.
The same question was given to 35 undergraduate students
in two different classes {\it immediately} after instruction in SGE. These students obtained $80\%$ on this question which is significantly better than the performance of
the graduate students before instruction in the graduate-level course. It appears that many of the first year graduate students enrolled in the graduate level quantum mechanics
course who took the survey had forgotten about the SGE. 
Moreover, discussions with some of the graduate students suggests that they had learned it only in the context of a modern physics course which was qualitative.

Question (2) in the Appendix 
is challenging because students have to realize that since the magnetic field gradient is in the $-x$ direction, the basis
must be chosen to be the eigenstates of $\hat S_x$ to readily analyze how the SGA will affect the spin state.
Here, the initial state, which is an
eigenstate of $\hat S_z$, $\vert \uparrow \rangle_z$, can be written as a linear superposition of the eigenstates of $\hat S_x$, i.e.,
$\vert \uparrow \rangle_z=(\vert \uparrow \rangle_x + \vert \downarrow \rangle_x) /\sqrt{2}$.
The magnetic field gradient in the $-x$ direction will couple the $\vert \uparrow \rangle_x$ and $\vert \downarrow \rangle_x$ 
components in the incoming spin state $\vert \uparrow \rangle_z$
with oppositely directed $x-$components of the linear momentum and will cause two spots on the phosphor screen separated along the $x$ axis.

Only $23\%$ of the more than 200 graduate students in a survey at the beginning of instruction provided the correct response. The performance of 35 undergraduate
students from two different classes who were given this question {\it immediately} after traditional instruction in SGE was only somewhat better ($39\%$).
Some undergraduate and graduate students were interviewed individually to better understand the reasoning behind their response.
In some of these interviews, we asked students to predict the outcome of these experiments and then showed them what actually
happens in a simulation and asked them to reconcile the differences between the observation and prediction.
This task turned out to be extremely difficult for students.
The most common difficulty in Question (2) was assuming that since the spin state is $\vert \uparrow \rangle_z$,
there should not be any splitting as shown in Figure 1.

Many students explained their reasoning by claiming that since the magnetic field gradient is in the $-x$ direction but
the spin state is along the $z$ direction, they are orthogonal to each other,
and therefore, there cannot be any splitting of the beam. Student responses suggest that
they were incorrectly connecting the gradient of the magnetic field in the 3D space
with the ``direction" of state vectors in the Hilbert space.  Several students in question (2) drew a monotonically increasing function.
Some of them incorrectly believed that the spin state in this situation will get pulled in one direction because the magnetic field gradient
is in a certain direction (see Figure 2). Asking the interviewed students explicitly about whether they could consider a basis that may be more 
appropriate to analyze this problem was rarely helpful.

One student drew the diagram shown in Figure 3 and described Larmor precession of spin but
did not mention anything about the spin-dependent momentum imparted to the particle due to the non-uniform magnetic field as in the SGE. 
Written responses and interviews suggest that many students were unclear about the fact that in a uniform external magnetic field, 
the spin will only precess 
(if not in a stationary state) but in a non-uniform magnetic field as in the SGE, there will be a spin-dependent momentum imparted to the particle
that may spatially separate the components of the spin angular momentum in the wave function under suitable conditions. 

\vspace*{-.22in}
\subsection{Larmor Precession of Spin involves Precession in Physical Space}
\vspace*{-.09in}

We note that the student who drew Figure 3 incorrectly believed that spin is due to motion in real space.
When he was reminded that the question was not about the dynamics (as suggested by the arrows drawn by the student
to show the direction of precession) but about the pattern observed on the screen,
he incorrectly claimed that the pattern on the screen would be a circle due to the precession of the spin in the magnetic field.
Similar to the difficulty
of this student, we have found that many students have difficulty realizing that spin is not an orbital degree of freedom and
we see two spots on the screen in questions (1) and (2) in the Appendix related to the SGE because
of the coupling of the spin degree of freedom with the orbital degree of freedom (e.g., the linear momentum).

\vspace*{-.22in}
\subsection{Difficulty with State Preparation}
\vspace*{-.09in}

The preparation of a specific quantum state may be challenging to achieve in the laboratory but it is relatively easy
to conceptualize theoretically at least in a 2D Hilbert space with SGE.
We find that the students have difficulty with the preparation of a specific quantum state even in a 2D Hilbert space.
Students were asked questions related to state preparation using SGA in both written tests and interviews, e.g., question (8) in
the Appendix.

A possible correct response would be to pass the initial beam through a SGA with a magnetic field 
gradient in the $x$ or $y$ direction and block one component of
the spatially separated beam that comes out of the SGA before passing it through another SGA with its field gradient in the $z$ direction.
One can then block the $\vert \uparrow \rangle_z$ component with a detector and obtain 
a beam in the spin state $\vert \downarrow \rangle_z$.

Out of 17 first year graduate students enrolled in quantum mechanics 
who had instruction in SGE, $82\%$ provided the correct response to question (8) in the Appendix. However, only $30\%$ of undergraduate students after
traditional instruction provided the correct response. Interviews suggest that students had great difficulty
thinking about how to choose an appropriate basis to facilitate the analysis of what should happen after particles in a given spin
state were sent through a SGA with a particular magnetic field gradient. 

\vspace*{-.22in}
\subsection{Difficulty in Differentiating Between a Superposition and a Mixture}
\vspace*{-.09in}

We also asked students to think of a strategy to distinguish between a superposition
in which all particles are in state $(\vert \uparrow \rangle_z + \vert \downarrow \rangle_z) /\sqrt{2}$
from a mixture in which half of the particles are in state $\vert \uparrow \rangle_z$ and the other half are in state $\vert \downarrow \rangle_z$
as in question (9) in the Appendix.

This question was very difficult for most students. One strategy for distinguishing between the superposition and the mixture given
is to pass each of them one at a time through a SGA with the field gradient in $-x$ direction. Then, since
$(\vert \uparrow \rangle_z + \vert \downarrow \rangle_z) /\sqrt{2}$ is $\vert \uparrow \rangle_x$, it will completely go out through the 
upper-channel after passing through a SGA with a negative $x$ gradient ($SGX_-$). On the other hand, the equal mixture of $\vert \uparrow \rangle_z$ and 
$\vert \downarrow \rangle_z$ will
have an equal probability of registering at the detectors in the lower and upper channels after the $SGX_-$ because these
states can be written as $(\vert \uparrow \rangle_x \pm \vert \downarrow \rangle_x) /\sqrt{2}$ in terms of the eigenstates of $\hat S_x$ and 
will become spatially separated after passing through the $SGX_-$.

Out of 17 first year graduate students enrolled in quantum mechanics 
who had instruction in SGE only $24\%$ provided the correct response to this question. 
In an undergraduate course in which the instructor
had discussed similar problems with students before giving them this question, $31\%$ provided the correct response after the traditional
instruction. 
One student incorrectly noted: ``Since the probability for an atom in the beam A to be in either state $\vert \uparrow \rangle_z$ or
$\vert \downarrow \rangle_z$ is 1/2, I can't distinguish it from B."
Another incorrect response emphasized differences in coupling of the spin angular momentum with the linear momentum: 
``The atoms in beam A will have their spin coupled to the $z$-component of their momentum. The other beams' atoms,
however, will not have $P_z$ coupled to $S_z$."
Some students who believed that it is possible to separate a mixture from
a superposition state using SGA provided incorrect reasoning. Figure 4 provides two such examples in which students first let each of the
beams pass through a SGA with a magnetic field gradient in the $z$ direction.

\vspace*{-.22in}
\section{SGE QuILT: Warm-up and Homework}
\vspace*{-.09in}

As discussed in the introductory section, the SGE QuILT builds on the prior knowledge of students and was developed based on the difficulties found via written surveys 
and interviews. 
The QuILT development went through a cyclical iterative process which includes the following stages:
(1) Development of the preliminary version based upon theoretical analysis of the underlying knowledge structure and research on students'
difficulties,
(2) Implementation and evaluation of the QuILT by administering it individually to students, measuring its impact on student learning and
assessing what difficulties remained,
(3) refinement and modification based upon the feedback from the implementation and evaluation.
When we found that the QuILT was working well in individual administration and the post-test performance
was significantly improved compared to the pre-test performance, it was administered in quantum mechanics classes. 

The SGE QuILT begins with a warm-up exercise and includes homework questions that students work on before and after working on the QuILT, respectively.
The warm-up exercise discusses preliminary issues such as why there is only a torque on the magnetic dipole in a uniform magnetic field but 
also a ``force" in a non-uniform magnetic field (or more precisely, a momentum is imparted to the particle due to its angular momentum as in the SGE). 
It also helps students understand that the divergence of the magnetic field
being zero according to the Maxwell's equation implies that the gradient of the magnetic field cannot be non-zero only in one direction and if we choose the gradient to be 
non-zero in two orthogonal directions and also apply a strong uniform magnetic field in one of those directions, the rapid Larmor precession will
make the average force in one of the directions zero. That way we can only focus on the magnetic field gradient in a particular direction
for determining its effect on the spin state after passing through the SGE. 

The warm-up exercise also discusses how the overall wavefunction of the quantum system includes both the spatial and spin parts of the wavefunction. 
For simplicity, students are asked to assume that before passing through a Stern Gerlach device with the field gradient in the $z$ direction (SGZ)
at time $t=0$, the spatial wave function $\psi(x,y,z)$ is a Gaussian localized near $(x,y,z)=(0,0,0)$
and the spatial and spin parts of the wave function are not entangled. Therefore, the overall wave function
which includes both the spatial and spin parts can be written as  $\Psi(t=0)=$ (orbital part) $\times$ (spin part), {\it i.e.,}
$\Psi(t=0)=\psi(x,y,z) \vert \chi \rangle$. Students are guided via a series of questions including the following:\\

{\it $\bullet$ A silver atom in the state $\Psi(t=0)=\psi(x,y,z) (a \vert \uparrow \rangle_z+b \vert \downarrow \rangle_z)$
passes through a SGZ with a non-uniform magnetic field
$\vec B=C_0 z \hat k$ from time $t=0$ to $t=T$. Which one of the following is the wave function at a time $t=T$
when the atom just exits the magnetic field? Assume that the atom is in the SGZ for a short time so that there is no change
in the $x,y,z$ coordinates. (Hint: The time development of each stationary state is via an appropriate
term of the type $e^{\pm i E_\pm t/\hbar}$.)\\
(a) $\Psi(T)=a \phi_+ \vert \uparrow \rangle_z+ b \phi_- \vert \downarrow \rangle_z$ where $\phi_\pm(x,y,z)=e^{\pm iC_0 \gamma z T/2} \psi(x,y,z)$ \\
(b) $\Psi(T)= \phi_+(x,y,z) (a \vert \uparrow \rangle_z+ b \vert \downarrow \rangle_z)$\\
(c) $\Psi(T)=\psi(x,y,z)(a \vert \uparrow \rangle_z+ b \vert \downarrow \rangle_z)$\\
(d) None of the above.}

Students further learn that in the wavefunction at time $T$, $\Psi(T)=a \phi_+ \vert \uparrow \rangle_z+ b \phi_- \vert \downarrow \rangle_z$
the spatial and spin parts of the wave functions are ``entangled" because spin and orbit cannot be factorized (i.e., cannot be
written in the form $\Psi(T)=\psi(x,y,z) \vert \chi \rangle$). Thus, measurement of the orbital degrees of freedom is linked to spin and vice versa.
Students are told that in the future discussion in the QuILT, the spatial part of the wave function $\psi(x,y,z)$ will not be mentioned explicitly. However,
they should understand that a SGA entangles the spatial and spin parts of the wave function.

The warm-up helps students understand how the coupling of the orbital and spin degrees of freedom causes the spatial separation 
of various spin components of the wave function. In the warm-up, students also learn that while the different components of spin may get
spatially separated after passing through a SGA, the wave function will remain in a superposition of different spin states until a measurement
is made, e.g., by placing a detector in an appropriate location. For example, the wave function for a spin-1/2 particle can
become spatially separated after passing through certain orientations of SGA and if a detector placed after the SGA at an appropriate location detects a particle (clicks), 
the wave function collapses to one state vs. when the detector does not click (in which case we have prepared the particles in a definite spin state).

In the SGE QuILT warm-up, students also learn about issues related to distinguishing between vectors in three-dimensional physical space and state vectors in Hilbert space. In this context, they learn that the 
magnetic field gradient in the $z$ direction is not perpendicular to a spin state in the Hilbert space, a common misconception among students. 
Students also learn about why choosing a particular basis is useful when analyzing particles going through a SGA with a particular magnetic field gradient. 
The SGE QuILT warm-up also helps clarify confusion about the $x$, $y$ and $z$ labels used to denote the orthogonal components of a vector, e.g., in classical mechanics, and the eigenstates of different 
components of spin operator ($\hat S_x$, $\hat S_y$  and $\hat S_z$) which are not orthogonal to each other. 

The SGE QuILT homework extends what students have learned in the tutorial and also focuses further on issues related to quantum measurement and state preparation via SGE. One common difficulty about SGE is that students often believe that a particle passing through a SGE is equivalent to the measurement of particle's spin angular momentum. These issues are clarified in the SGE QuILT homework.

\vspace*{-.22in}
\section{SGE QuILT}
\vspace*{-.09in}

As noted earlier, the SGE QuILT uses a guided inquiry-based approach in which various concepts build on each other gradually.
It employs visualization tools to help students build a physical intuition about concepts related to the SGE. The
Open Source Physics SPINS program~\cite{mario} was adapted as needed for the SGE QuILT which
extends David McIntyre's open source Java applet~\cite{mcintyre} by allowing simulated experiments to be stored and run easily. 
One effective strategy to help students build a robust knowledge structure is by
causing a cognitive conflict in students' minds such that the students themselves realize that there
is some inconsistency in their reasoning and then providing them appropriate guidance and support.
In the SGE QuILT, after predicting what they expect in
various situations, students are asked to check their predictions using simulations. If the prediction and observations
don't match, students reach a state of cognitive conflict. At that point the QuILT provides them guidance to
help build a good grasp of relevant concepts and reconcile the difference between their predictions and observations. 

As noted earlier, the SGE QuILT helps students learn about issues related to measurement, preparation of a desired quantum state, e.g.,
$\vert \uparrow \rangle_x$, starting with an arbitrary initial state, time-development of the wave function,
the difference between superposition and mixture, the difference between physical space and Hilbert space, the importance of choosing an appropriate basis
to analyze what should happen in a particular situation, etc. 
Figure 6 shows a simulation constructed from the OSP SPINS~\cite{mario} program that students work with after their initial prediction related to a question
that shows that one can input
$\vert \uparrow \rangle_z$ and obtain $\vert \downarrow \rangle_z$. Students again have to reconcile the difference between their prediction
and observation with suitable hints.

In order to help students understand that it is possible to input $\vert \uparrow \rangle_z$ through the SGAs
and prepare an orthogonal state $\vert \downarrow \rangle_z$ on the way out, the QuILT also draws an analogy with the photon polarization
states. Students learn that if atoms in the state $\vert \uparrow \rangle_z$ pass through a SGZ only, 
the state $\vert \downarrow \rangle_z$ will not be obtained on the way out. However, $\vert \downarrow \rangle_z$ is obtained
in the simulated experiment in Figure 6 because we have inserted $SGX_-$ (a SGA with the field gradient in the negative x-direction) at an intermediate stage. Students consider the analogy with
vertically polarized light passing directly through a horizontal polarizer (Figure 7 a) vs.
passing first through a polarizer at $45^0$ followed by a horizontal polarizer (Figure 7 b).
There is no light at the output if vertically polarized light passes directly through a horizontal polarizer.
On the other hand, if the polarizer at $45^0$ is present, light becomes polarized at
$45^0$ after the $45^0$ polarizer, which is a linear superposition of horizontal and vertical polarization.
Therefore, some light comes out through the horizontal polarizer placed after the $45^0$ polarizer. 
Since the experiment with the polarizers (in the context of a photon beam not a single photon) 
is familiar to students from introductory physics, this analogy can help students learn about the SGE using a familiar context.

While working through the QuILT, students are asked a guided sequence of questions to help them distinguish between superposition and mixture. 
The QuILT presents a common incorrect point of view on the issue dealing with superposition and mixture. 
Then, the students are given an opportunity to check their predictions using computer simulations and reconcile the differences using more guidance 
and support as needed.  Further questions are given to students to help them understand the difference between a pure state and a mixture 
by reinforcing the analogy between the spin states of electrons and the polarization states of photons.
The guidance provided to students is decreased as students make progress through the QuILT. In the later part of the QuILT, students are given open-ended questions such
as the following:

{\it The following questions relate to the simulation ``Unknown State". Run the simulation "Unknown State" first. Then, answer the following questions.

$\bullet$ Write down at least 3 different possible spin states of the incoming particles that will show the behavior seen in the simulation. 
The incoming particles need not necessarily have identical spin states (can be a mixture). Explain your reasoning for your choices.

$\bullet$ Choose two of the different possible spin states you predicted for the simulation you saw. 
Now come up with some simulations using SGAs that would distinguish between these two possible spin states. 
You can choose one or more SGAs to find out which of the two spin states it is. Share your set-up with others in your class.}

\vspace*{-.22in}
\section{Pre-test and post-test data for SGE QuILT}
\vspace*{-.09in}

We conducted preliminary evaluations of the SGE QuILT in two junior-senior level classes, first with 22 and second with 13 undergraduate students. 
The two classes were taught by different instructors. 
In both classes, students first received traditional instruction about the SGE, took a pre-test, worked on the tutorial and then took
a post-test in the following class period. The test questions are given in the Appendix. 
In particular, the first class with 22 students was given questions (1)-(4) in the pre-test and questions (5)-(7) on the post-test.
The average pre-test score for this class was $53\%$ and the average post-test score was $92\%$.

For the second class, we designed two versions of a test (versions A and B) to assess student learning. Version A contained questions (1), (2), (3), (4) and (9) while 
version B had questions (1), (2), (5), (6),and (7) (see the Appendix for a description of all the questions).  
Students in the second class were randomly administered either version A or version B of the test as the pre-test after the traditional instruction.
Then, each student was administered the version of the test he/she had not taken as the post-test after working on the QuILT.
In particular, 8 students in that class were administered version A as pre-test  (and version B as the post-test) whereas the
other 8 students were given version B as the pre-test (and version A as the post-test).
The average pre-test score for this second class was $37 \%$ and the average post-test score was $84\%$. The average pre-test and post-test performance on each question combining the 
two groups of students is given in Table 1. Except for Question (1), on which students performed reasonably well even on the pre-test (after traditional instruction), student 
performance improved on all the other questions after working on the QuILT.

In Table 1, the improved performance on question (2) (in which students were asked about the pattern on the screen when neutral silver atoms in the spin state $\vert \uparrow \rangle_z$
were sent through a $SGX_-$) after the QuILT suggests that students were much more likely to be able to predict the type of pattern that
should form on the screen when particles in a particular spin-state pass through a SGA with a particular field gradient.
Individual discussions with some students suggest that after the QuILT students 
had a reasonably good understanding of how to choose a good basis to analyze the spin state of a particle passing through a SGA with a particular field gradient.
Some of them were not only able to write the initial spin state in an appropriate basis, 
they were able to differentiate between the spin states which are vectors in the Hilbert space and the direction of the magnetic field gradient in the physical space because these are vectors
in different spaces. 
In particular, during the discussions, some students correctly noted explicitly that the eigenstates of the $z$-component of spin are orthogonal to each other but not orthogonal to the magnetic field 
gradient in the $z$ direction in physical space. 
In question (3), many students realized after the QuILT that the given superposition of the eigenstates of the $z$-component of spin is actually an eigenstate of the x-component of spin. 
Student performance after the QuILT on question (7) (in which the incoming state was a general state) further suggests that they had a better understanding of how to choose a convenient basis
to analyse the output of a SGA than before the QuILT.
Students also performed reasonably well after the QuILT on questions where the particle went through several SGAs in tandem (e.g., quetions (4) and (6)).
The improved performance on questions (5) and (9) (in which question (9) was open-ended)
suggest that students had a better understanding of how a superposition of spin states and a mixture can be differentiated using SGAs.
Furthermore, the improvement in the open-ended question about the preparation of a particular spin state starting from another spin-state using a SGA in question (8) is encouraging.

In addition to the pre-test and post-test, students who had used the SGE QuILT were asked the following two questions after five
months in the second semester junior-senior level undergraduate quantum mechanics course. The goal was to investigate whether
students can distinguish the two situations, one of which involves a superposition and another a mixture when the magnetic field 
gradient was explicitly provided (this question is somewhat different from question (8) on the post-test given to students five months ago
in which students had to come up with their own arrangement of the SGAs):

{\it (a) Suppose a beam consists of silver atoms in the state $(\vert \uparrow \rangle_z+\vert \downarrow \rangle_z)/\sqrt{2}$.
The beam passes through a Stern Gerlach apparatus (SGA) with the magnetic field gradient in the \underline{x-direction}. 
How many detector(s) are sufficient to detect all the silver atoms passing through the SGA?
Draw a diagram and explain your reasoning.

(b) Suppose a beam consists of an unpolarized mixture of silver atoms in which half of the silver atoms are in state
$\vert \uparrow \rangle_z$ and half are in state $\vert \downarrow \rangle_z$.
The beam passes through a SGA with the magnetic field gradient in the \underline{x-direction}. 
How many detector(s) are sufficient to detect all the silver atoms passing through the SGA?
Draw a diagram and explain your reasoning.
}

Eight out of nine undergraduate students who answered these two questions at the end of the second semester provided the correct response for both questions.
It is encouraging that the students had retained these concepts a full semester after working on the QuILT.

\vspace*{-.22in}
\section{Summary}
\vspace*{-.07in}

We have investigated students' difficulties in quantum mechanics via the SGE and used the findings as a guide to develop a SGE QuILT.
The Stern-Gerlach experiment can be used to teach many aspects of quantum mechanics
effectively including issues related to measurement, importance of choosing a particular
basis, differentiation between Hilbert space and real space, and
the difference between a pure linear superposition of states vs. a mixture.
Preliminary evaluation suggests that the QuILT is effective in improving students' understanding of concepts related to SGE.

\vspace*{-.23in}
\section{Acknowledgments}
We thank the instructors for administering the surveys and QuILTs in their classes.
We are grateful to NSF for awards PHY-0653129 and 055434.

\begin{table}[h]
\centering
\begin{tabular}[t]{|c|c|c|}
\hline
Question&Pre-test Score in $\%$ (Number of Students)& Post-test Score in $\%$ (Number of students)\\[0.5 ex]
\hline
1&80 (35)&81 (13)\\[0.5 ex]
\hline
2&39 (35)&77 (13)\\[0.5 ex]
\hline
3&34 (30)&80 (5)\\[0.5 ex]
\hline
4&47 (30)&80 (5)\\[0.5 ex]
\hline
5&60 (5)&94 (30)\\[0.5 ex]
\hline
6&0 (5)&92 (30)\\[0.5 ex]
\hline
7&0 (5)&92 (30)\\[0.5 ex]
\hline
8&30 (5)&100 (8)\\[0.5 ex]
\hline
9&31 (8)&70 (5)\\[0.5 ex]
\hline
\end{tabular}
\vspace{0.1in}
\caption{The pre-test (after traditional instruction but before the QuILT) and post-test (after the QuILT) scores on each question. The total number of students including both classes who answered
each question is given in parenthesis. Each student in a class of 22 students was given the same pre-test (similarly, all the 22 students received the same post-test). The pre-test
and post-test were mixed for the second
class of 13 students as discussed in the text
(in particular, 5 students out of 13 answered as post-test questions what the other 8 students answered as pre-test questions). }
\label{junk2}
\end{table}

\pagebreak

{\bf Appendix: The Pre-/Post-test Questions}

Note: Some of the questions below (or questions similar to them) were also used during the investigation of students' difficulties at various stages of the development of the QuILT.

The following information is provided in the pre-/post-test.

The following pictorial representations are used for a Stern-Gerlach apparatus (SGA). 
\begin{center}
\epsfig{file=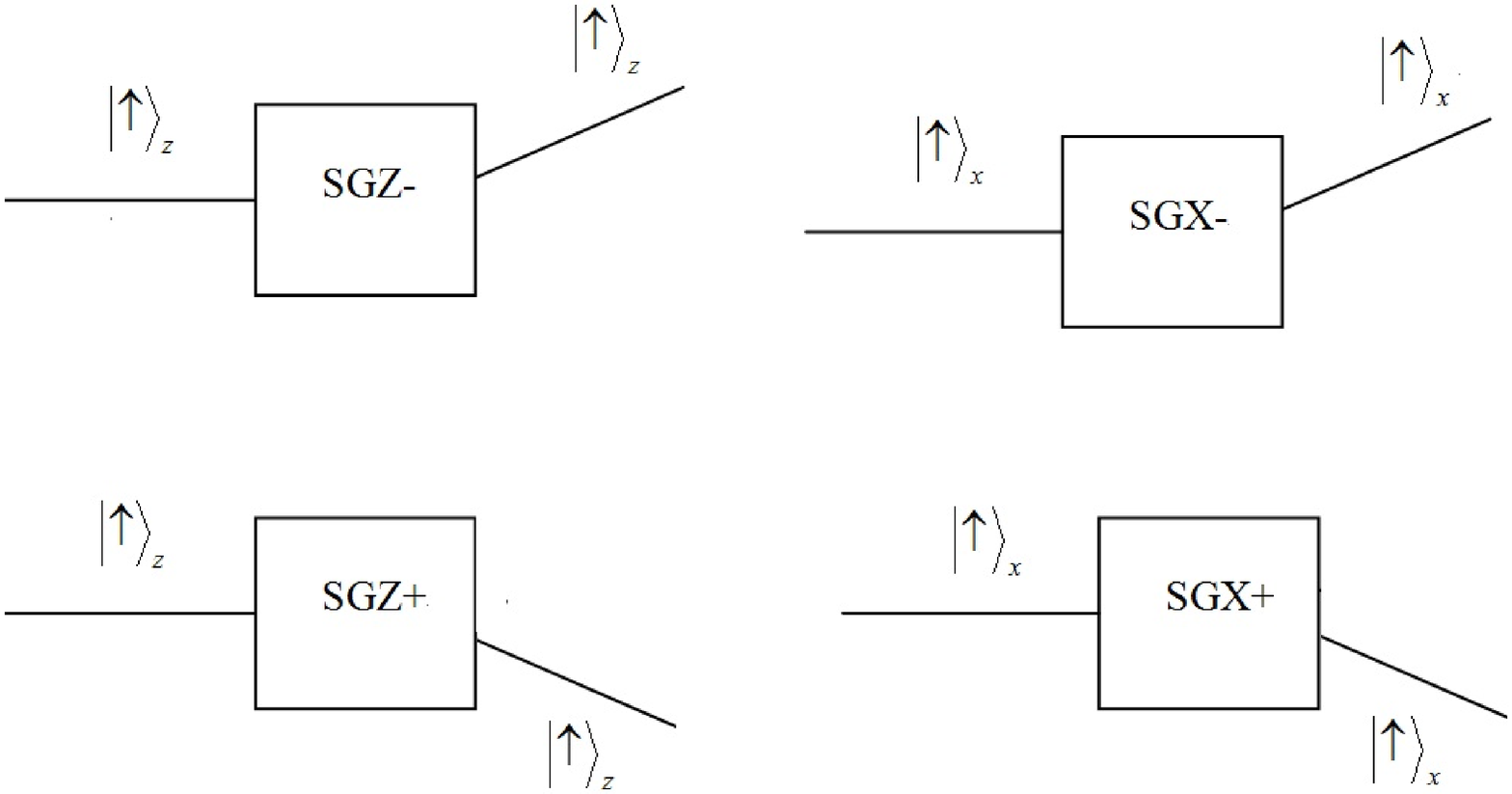,height=3.2in}
\end{center}
If an atom in state $\vert \uparrow \rangle_z$ (or $\vert \downarrow \rangle_z$) 
passes through a SGA with the field gradient in the negative z direction ($SGZ_-$), it will be deflected in the +z (or -z) direction. 
If an atom in state $\vert \uparrow \rangle_z$ (or $\vert \downarrow \rangle_z$)
passes through a SGA with the field gradient in the positive z direction ($SGZ_+$), 
it will be deflected in the -z (or +z) direction. Similarly, if an atom in state $\vert \uparrow \rangle_x$  passes through a $SGX_-$ (or $SGX_+$), 
it will be deflected in the +x (or -x) direction. The figures above show examples of deflections through the SGX and SGZ 
in the plane of the paper. However, note that the deflection through a SGX will be in a plane perpendicular to the deflection through a SGZ. 
This actual three-dimensional nature should be kept in mind in answering the questions.

\noindent
Notation: $\vert \uparrow \rangle_z$ and  $\vert \downarrow \rangle_z$ represent the orthonormal eigenstates of
$\hat S_z$ (the $z$ component of the spin angular momentum).
SGA is an abbreviation for a Stern-Gerlach apparatus.\\

\noindent
(1) A beam of neutral silver atoms propagating along the $y$ direction (into the page) in spin state
$(\vert \uparrow \rangle_z+\vert \downarrow \rangle_z)/\sqrt{2}$
is sent through a SGA with a vertical magnetic field gradient in the $-z$ direction.
Sketch the pattern you expect to observe on a distant phosphor screen in the x-z plane when the atoms hit the screen. Explain your reasoning.

\begin{center}
\epsfig{file=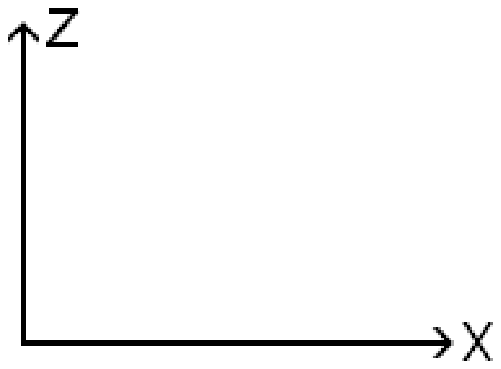,height=1.in}
\end{center}

\noindent
(2) A beam of neutral silver atoms propagating along the $y$ direction (into the page) in spin state $\vert \uparrow \rangle_z$
is sent through a SGA with a horizontal magnetic field gradient in the $-x$ direction.
Sketch the pattern you expect to observe on a distant phosphor screen in the x-z plane when the atoms hit the screen. Explain your reasoning.

\begin{center}
\epsfig{file=coordinate_z_x.eps,height=1.in}
\end{center}

\noindent
(3) Chris sends silver atoms in an initial spin state $\vert \chi(0)  \rangle = (\vert \uparrow \rangle_z+ 
\vert \downarrow \rangle_z)/\sqrt{2}$
one at a time through a $SGX_-$. He places a ``down" detector in appropriate location as shown. What is the probability of the detector 
clicking when an atom exits the $SGX_-$?
\begin{center}
\epsfig{file=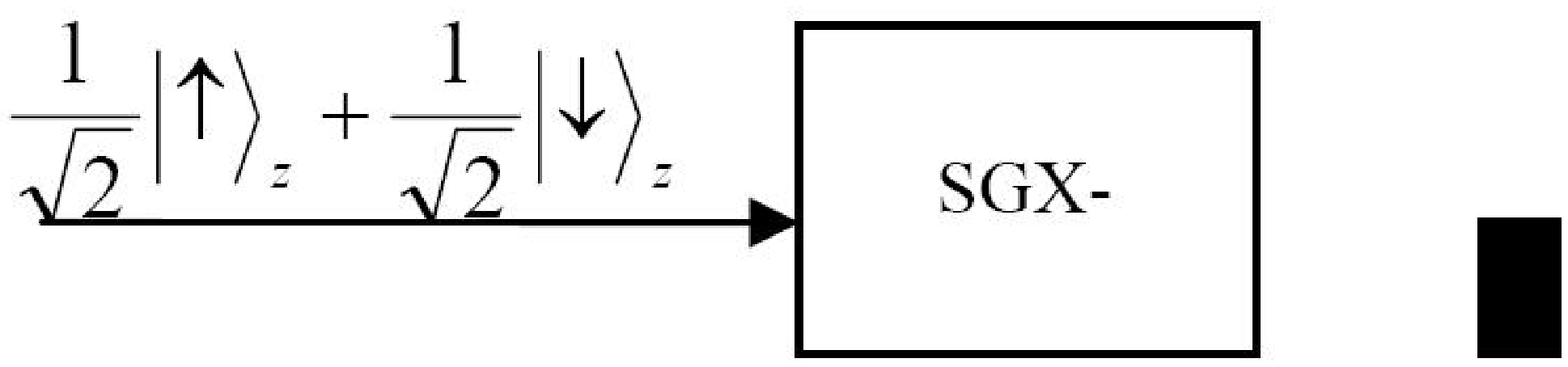,height=.76in}
\end{center}

\noindent
(4) Silver atoms in an initial spin state $\vert \chi(0)  \rangle = \vert \uparrow \rangle_z$ pass one at a time
through two SGAs with the magnetic field gradients as shown below. Two suitable detectors are placed, one after the
first SGA and the second at the end to detect the atoms after they pass through both SGAs. The atoms that do not
register in the ``up" detector at the end are collected for another experiment. Find the fraction of atoms that
are detected in the ``up" detector at the end and the normalized spin state of the atoms that are collected for another experiment.
\vspace{-0.05in}
\begin{center}
\epsfig{file=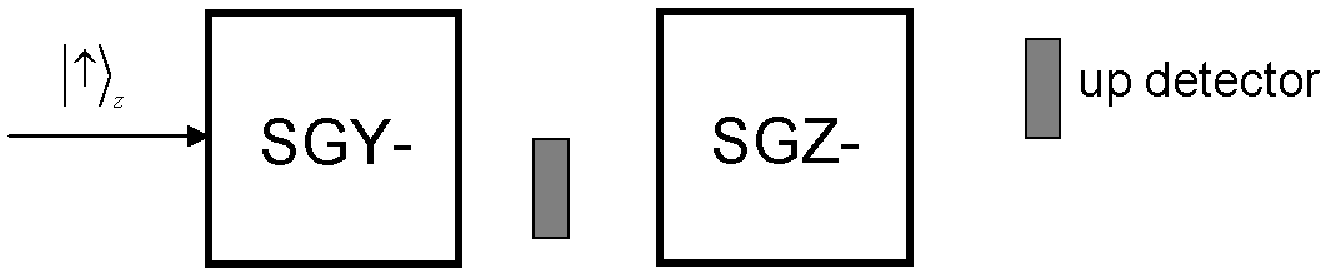,height=.8in}
\end{center}
\vspace{-0.05in}

\noindent
(5) Suppose beam A consists of silver atoms in the state 
$\vert \chi(0) \rangle = (\vert \uparrow \rangle_z+ \vert \downarrow \rangle_z)/\sqrt{2}$, and beam B is an unpolarized mixture in which half of 
the silver atoms are in state $\vert \uparrow \rangle_z$ and half are in state $\vert \uparrow \rangle_z$. 
Choose all of the following statements that are correct.\\
\noindent
\hspace*{0.45in} (1) Beam A will not separate after passing
through SGZ (either $SGZ_-$ or $SGZ_+$).\\
\hspace*{0.45in} (2) Beam B will split into two parts after 
passing through SGZ.\\
\hspace*{0.45in} (3) We can distinguish between beams A and 
B by passing each of them through SGX.\\
A. (1) only\\
B. (2) only\\
C. (1) and (2) only\\
D. (2) and (3) only\\
E. All of the above

\noindent
(6) Sally sends silver atoms in state $\vert \uparrow \rangle_z$ through three SGAs as shown.
Next to each detector, write down the probability that the detector clicks. 
The probability for the clicking of a detector refers to the probability that a particle entering the {\bf first}
SGA reaches that detector.
Also, after each SGA, write the spin state Sally has prepared. Explain.
\begin{center}
\epsfig{file=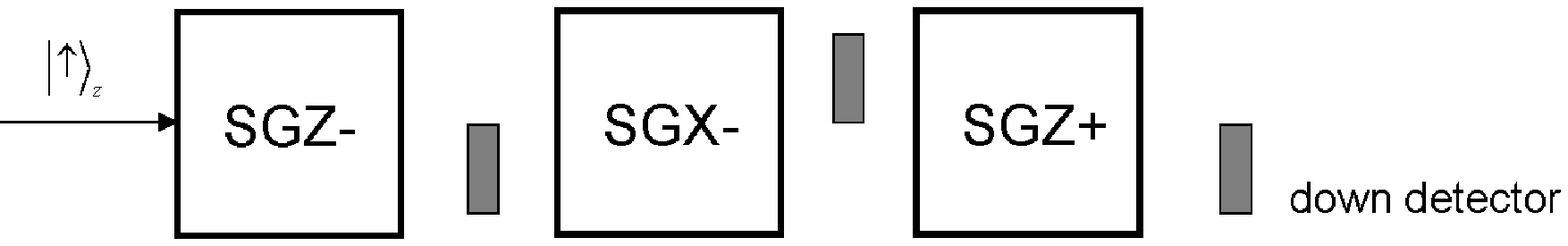,height=0.8in}
\end{center}

\noindent
(7) Harry sends silver atoms all in the normalized spin state $\vert \chi \rangle =a \vert \uparrow \rangle_z +
b \vert \downarrow \rangle_z$ through an $SGX_-$.
He places an ``up" detector as shown to block some silver atoms and collects the atoms coming out in the ``lower channel"
for a second experiment. What fraction of the initial silver atoms will be available for his second experiment? What is the
spin state prepared for the second experiment? Show your work.
\begin{center}
\epsfig{file=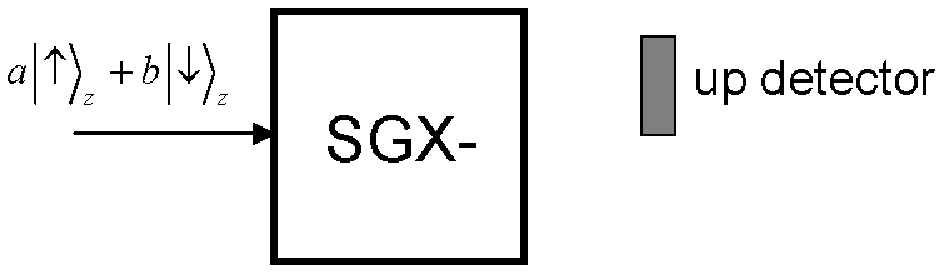,height=.8in}
\end{center}

\noindent
(8) Suppose you have a beam in the spin state $\vert \chi(0) \rangle=\vert \downarrow \rangle_z$ but you need to prepare the spin state
$\vert \uparrow \rangle_z$ for your experiment. Could you use Stern-Gerlach Apparati and detectors to prepare the spin state 
$\vert \uparrow \rangle_z$? If yes, sketch your setup below and explain how it works. If not, explain why not.

\noindent
(9) Suppose beam A consists of silver atoms in the state $(\vert \uparrow \rangle_z + \vert \downarrow \rangle_z) /\sqrt{2}$, 
and beam B consists of an unpolarized mixture in which half of the silver atoms 
are in state $\vert \uparrow \rangle_z$ and half are in state $\vert \downarrow \rangle_z$. 
Design an experiment with Stern-Gerlach Apparati and detectors to differentiate these two beams. 
Sketch your experimental setup below and explain how it works. 

\pagebreak

\begin{figure}
\epsfig{file=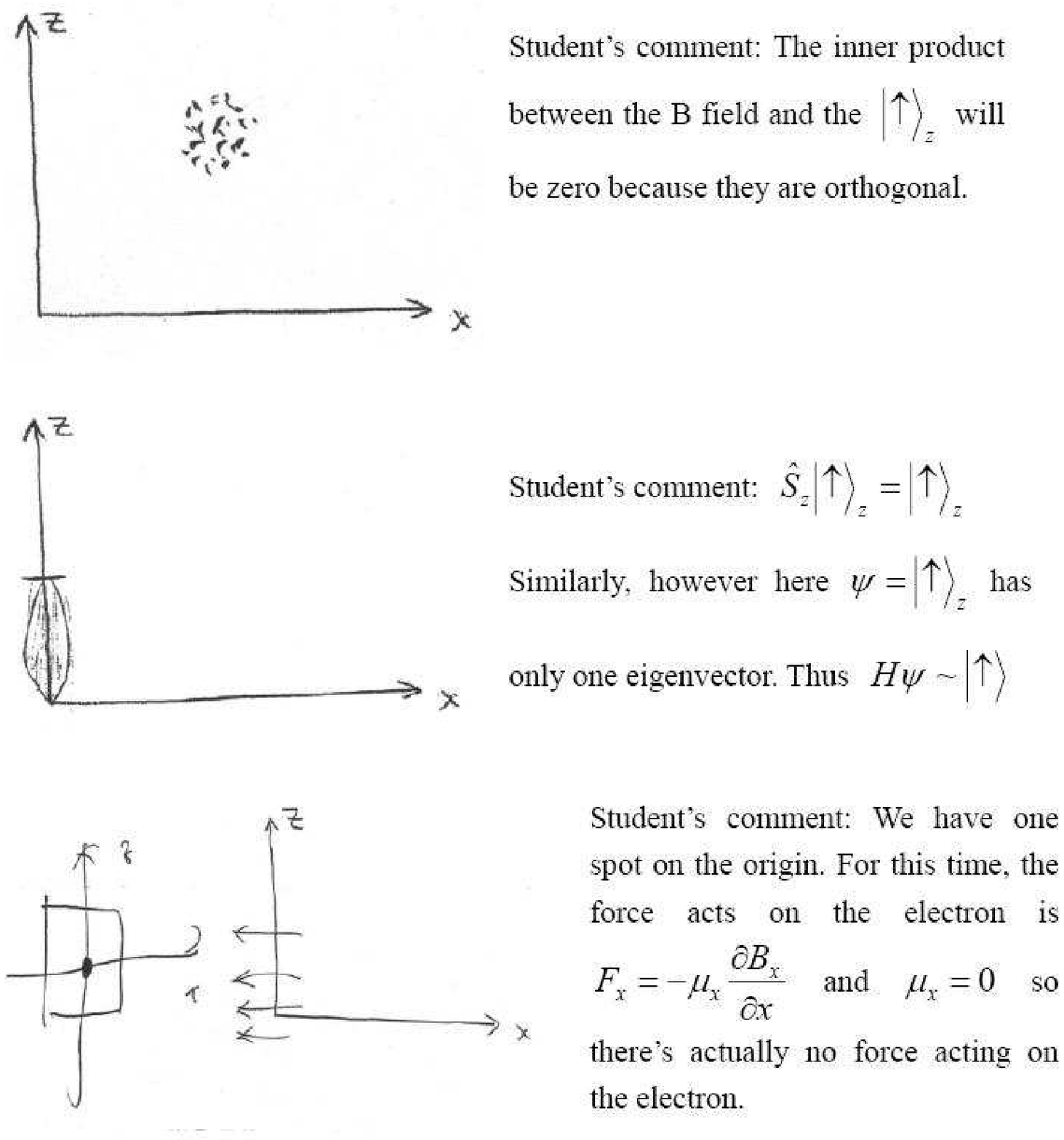,height=4.46in}
\caption{Three sample responses in which students provided incorrect explanations for why there should be one spot instead of two in question (2) in the Appendix. The students' comments with each figure are typed for clarity.}
\end{figure}

\pagebreak

\begin{figure}
\epsfig{file=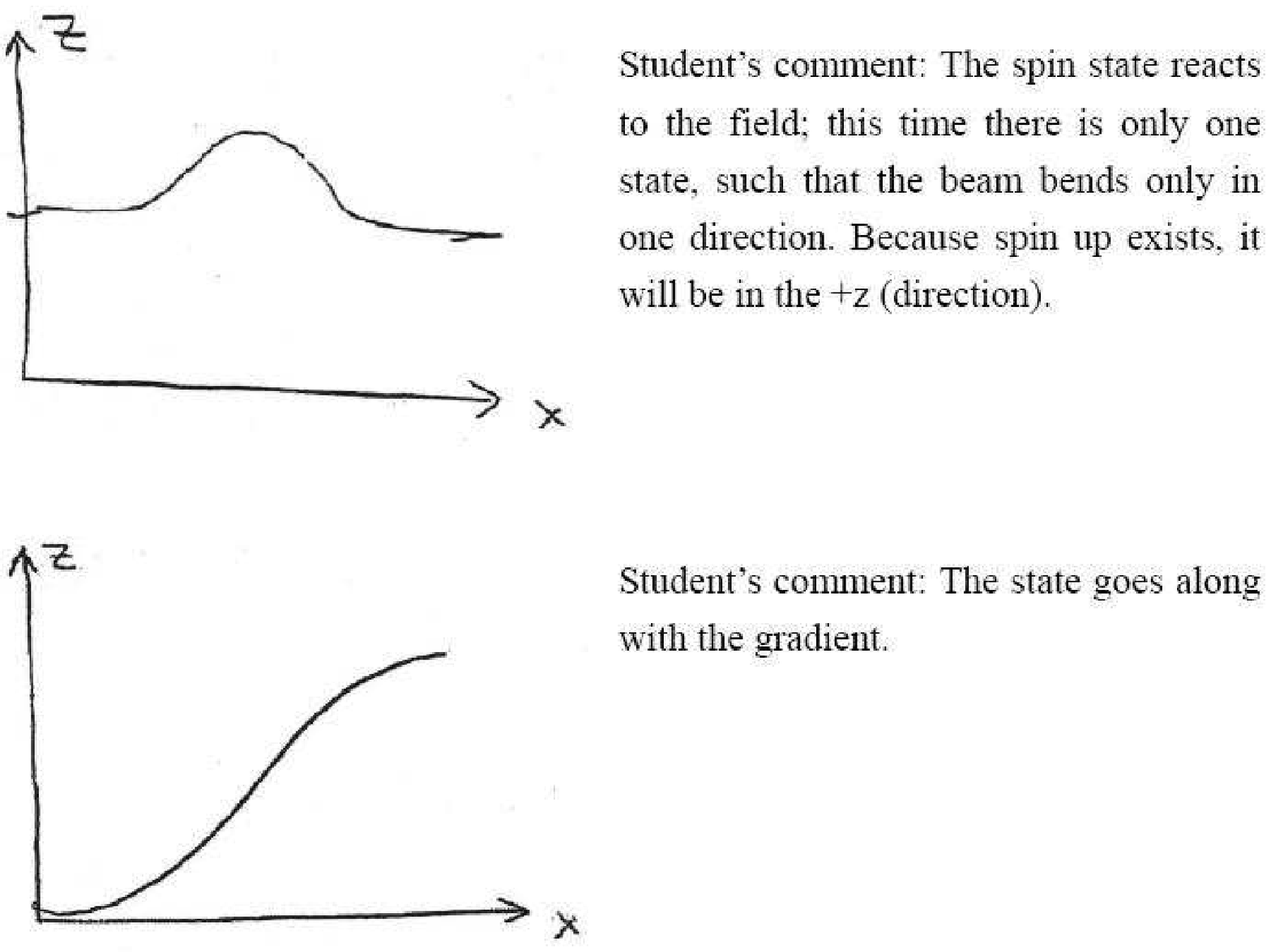,height=3.9in}
\caption{Two sample responses in which students provided incorrect explanations for why the state/beam will bend as shown in response
to the magnetic field gradient in question (2) in the Appendix. The students' comments with each figure are typed for clarity.}
\end{figure}

\pagebreak

\pagebreak

\begin{figure}
\epsfig{file=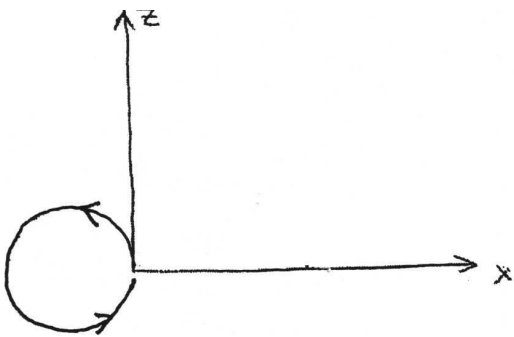,height=1.716in}
\caption{A diagram drawn by a student showing the Larmor precession of spin in response to question (2) in the Appendix. }
\end{figure}

\pagebreak

\begin{figure}
\epsfig{file=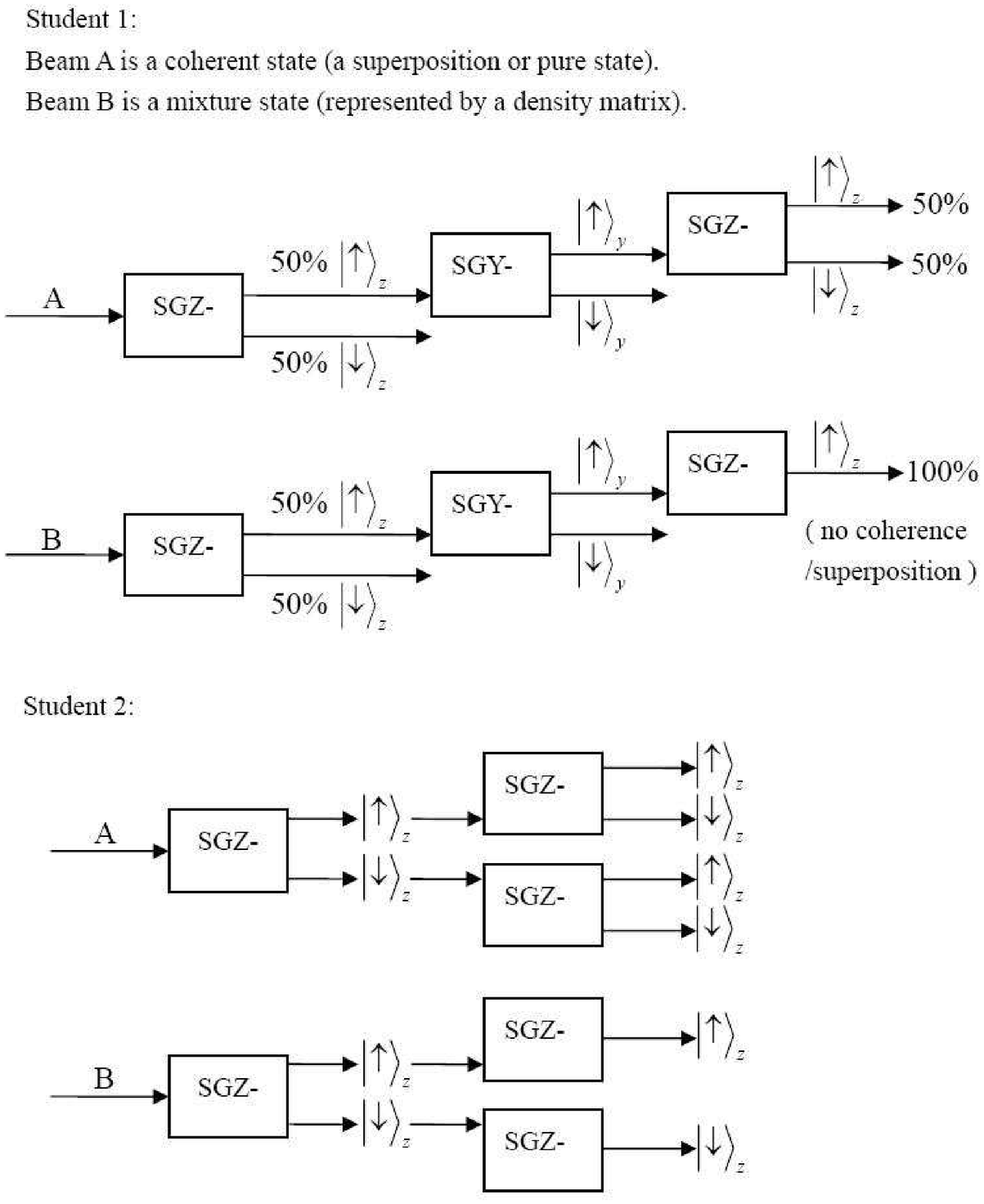,height=5.8in}
\caption{Examples of two graduate students' responses to question (9). The students' responses are typed for clarity.}
\end{figure}

\pagebreak
\begin{figure}
\epsfig{file=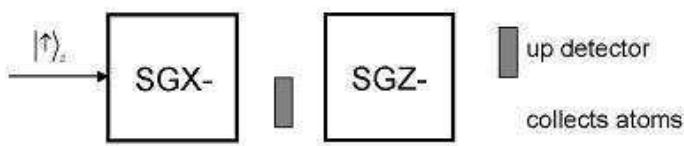,height=0.9in}
\caption{Set up for a guided example in the QuILT.}
\end{figure}

\pagebreak

\begin{figure}
\epsfig{file=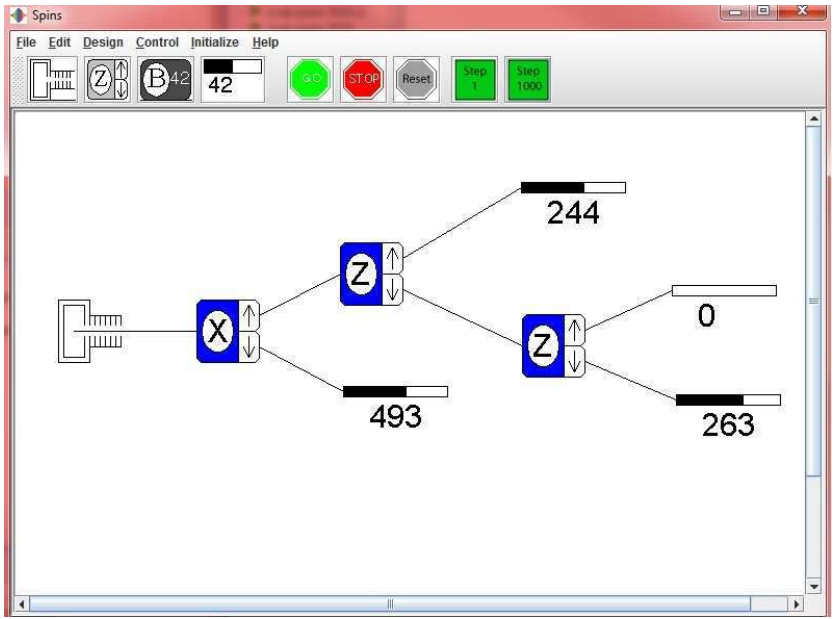,height=3.2in}
\caption{A snapshot of the simulated experiment constructed from the OSP SPINS program~\cite{mario} that students play with that shows that one can input
$\vert \uparrow \rangle_z$ and obtain $\vert \downarrow \rangle_z$. This snapshot shows 493 particles are registered in the detector right after passing
through the SGA with the magnetic field gradient in $x$ direction ($SGX_-$), 244 particles are registered in the detector right after the first SGA with the magnetic 
field gradient in the $z$ direction ($SGZ_-$) and 263 particles are registered in the detector after the second $SGZ_-$.}
\end{figure}

\pagebreak

\begin{figure}
\epsfig{file=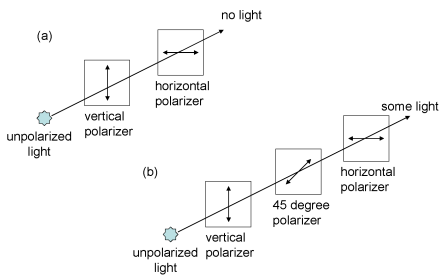,height=2.9in}
\caption{Analogy between spin states and photon polarization states.}
\end{figure}

\end{document}